\documentclass[conference]{IEEEtran}
\IEEEoverridecommandlockouts
\usepackage{cite}
\usepackage{subcaption}
\usepackage{amsmath,amssymb,amsfonts}
\usepackage{algorithmic}
\usepackage{graphicx}
\usepackage{textcomp}
\usepackage{xcolor}
\usepackage{cleveref}
\def\BibTeX{{\rm B\kern-.05em{\sc i\kern-.025em b}\kern-.08em
    T\kern-.1667em\lower.7ex\hbox{E}\kern-.125emX}}
\begin{document}

\title{Strategies and Challenges of Efficient White-Box Training for Human Activity Recognition\\
}

\author{\IEEEauthorblockN{Daniel Geißler}
\IEEEauthorblockA{\textit{German Research Center for} \\\textit{Artificial Intelligence (DFKI)}\\
Kaiserslautern, Germany\\
daniel.geissler@dfki.de}
\and
\IEEEauthorblockN{Bo Zhou}
\IEEEauthorblockA{\textit{German Research Center for} \\\textit{Artificial Intelligence (DFKI)}\\
Kaiserslautern, Germany\\
bo.zhou@dfki.de}
\and
\IEEEauthorblockN{Paul Lukowicz}
\IEEEauthorblockA{\textit{German Research Center for} \\\textit{Artificial Intelligence (DFKI)}\\
Kaiserslautern, Germany\\
paul.lukowicz@dfki.de}

}

\maketitle

\begin{abstract}
Human Activity Recognition using time-series data from wearable sensors poses unique challenges due to complex temporal dependencies, sensor noise, placement variability, and diverse human behaviors. 
These factors, combined with the nontransparent nature of black-box Machine Learning models impede interpretability and hinder human comprehension of model behavior.
This paper addresses these challenges by exploring strategies to enhance interpretability through white-box approaches, which provide actionable insights into latent space dynamics and model behavior during training. 
By leveraging human intuition and expertise, the proposed framework improves explainability, fosters trust, and promotes transparent Human Activity Recognition systems.
A key contribution is the proposal of a Human-in-the-Loop framework that enables dynamic user interaction with models, facilitating iterative refinements to enhance performance and efficiency. 
Additionally, we investigate the usefulness of Large Language Model as an assistance to provide users with guidance for interpreting visualizations, diagnosing issues, and optimizing workflows.
Together, these contributions present a scalable and efficient framework for developing interpretable and accessible Human Activity Recognition systems.
\end{abstract}


\section{Introduction}
Human Activity Recognition (HAR) focuses on identifying and classifying human behaviors based on sensor-generated time-series data \cite{lara2012survey}. 
This field holds significant promise for applications such as healthcare monitoring, personalized fitness tracking, and smart environments \cite{wang2019deep}. 
Especially for wearable sensor inputs, for instance, Inertial Measurement Units \cite{kwapisz2011activity}, Impedance \cite{liu2024imove} or Capacitance \cite{geissler2024embedding}, the complexities inherent in those time-series data pose substantial challenges. 
Temporal dependencies, overlapping activity patterns, and variability in sensor quality and placement contribute to difficulties in model training and evaluation. 
These factors often result in suboptimal and inefficient performance, with traditional black-box Machine Learning (ML) models failing to provide the necessary transparency for diagnosing and addressing these issues \cite{lipton2018mythos}.

The lack of interpretability in black-box models hinders not only the understanding of their decision-making processes but also their usability in critical applications \cite{guidotti2018survey}. 
For instance, when models misclassify activities due to overlapping features or noisy inputs, practitioners often cannot pinpoint the root cause \cite{ribeiro2016should}. 
This opacity complicates optimization efforts, leading to inefficiencies in training and unnecessary consumption of computational resources for investigation and resolving \cite{geissler2024power}.
Furthermore, this creates a gap in trust, particularly in domains where understanding model behavior is crucial \cite{doshi2017towards}.

To address these limitations, our work presents a framework that transitions from black-box to white-box training for HAR models. 
The cornerstone of this approach is the utilization of visualization metrics to reveal latent space dynamics during black-box-based model training. 
These visual representations provide insights into how the model processes and learns from time-series data, enabling practitioners to identify issues such as poor feature separability or learning inconsistencies. 
By making the training process more transparent, we empower users to actively contribute to optimizing the model and shrink its required resources.

Building on this, we propose a Human-in-the-Loop (HITL) approach that allows for real-time user interaction with the training process. 
By integrating human intuition into the workflow, the HITL framework facilitates targeted interventions, improving both the model’s performance and the efficiency of the training process. 
Additionally, we envision the support of Large Language Models (LLMs) as smart assistance to support users in interpreting the visualizations, diagnosing potential issues, and suggesting corrective measures. 
This synergistic combination of human intuition, visual insights, and LLM guidance aims to bridge the gap between complex model behavior and user understanding.

\section{Related Work}
\label{sec:related}
Interactive visualizations and explainability techniques have become essential tools in the development of interpretable ML models. 
These approaches aim to improve model transparency, enhance user trust, and aid in the refinement of ML systems through human interaction and insight.

\subsection{Interactive Machine Learning Visualizations}

Interactive visualizations are powerful tools that can unveil the internal workings of ML models, helping users understand and improve their behavior through dynamic and engaging interfaces. 
By enabling users to interrogate, explain, and validate the decision-making process, these tools help open up the black-box nature of ML models \cite{Hurley2021Interactive}
Various tools are available, essentially aiming to enhance trust in ML models through increasing human perception of the model's internals \cite{geissler2023latent}.
TensorFlow Graph Visualizer for instance helps users understand complex ML architectures by visualizing their underlying dataflow graphs, making it easier to debug and improve model structures \cite{Wongsuphasawat2018Visualizing}.
However, it is still a challenge to select the right visualization, especially when lossy dimension reduction strategies are applied and thus information content is missing \cite{Rauber2017Visualizing}.

Further, the interactivity of visualizations can elevate the insights and feed the user's input back into the model for adaptation \cite{Fuchs2009Visual}.
Works like \cite{Sacha2017What} have proven the advantages of Human-centered and HITL systems to enhance ML development through human knowledge integration and feedback.
Similar to the idea of this work, \cite{wei2022fine} proposes a human-in-the-loop approach to enhance deep neural network classification accuracy by leveraging human knowledge. 
By projecting high-dimensional latent spaces onto a two-dimensional workspace, users can interactively modify coordinates, which are then fed back to fine-tune the network, improving prediction performance.

\subsection{Explainability and Interpretability for Deep Learning Latent Representations}




Explainability and interpretability are essential for understanding model behavior, fostering trust, and improving performance. 
In the context of deep learning, disentangled and interpretable representations have been a central focus. 
Generative Adversarial Networks (GANs) have been effectively applied to achieve disentanglement learning by maximizing mutual information, enabling tasks such as attribute manipulation and image editing \cite{chen2016infogan, jeon2021ib}. 
Additionally, Beta-variational autoencoders ($\beta$-VAEs) promote sparse latent representations that encourage disentanglement, improving model interpretability \cite{burgess2018understanding}. 
Latent quantization has also emerged as a method to further disentangle the representations, improving the interpretability of deep learning models by quantizing latent space and achieving better feature separation \cite{hsu2024disentanglement}.

High-dimensional data poses challenges for interpretability, often requiring dimensionality reduction while retaining human interpretability. Frameworks leveraging semantically meaningful latent representations and adapting the Shapley paradigm offer efficient, model-agnostic explanations \cite{de2020human}. Similarly, invertible interpretation networks provide a transparent view of hidden representations without retraining, facilitating a better understanding of complex latent variables \cite{esser2020disentangling}.

A significant challenge in model interpretability lies in prototype-based learning, where decisions are based on similarities to learned latent prototypes. The semantic gap between latent space similarity and input space similarity can lead to issues in interpretability, often confusing decision-making. Addressing this gap is crucial for ensuring that prototype-based models remain transparent \cite{hoffmann2021looks}.

Several comprehensive surveys have also been published, providing overviews of interpretability methods in deep learning. 
These surveys clarify the definitions of interpretability, propose taxonomies, and evaluate various techniques for model explainability. They stress the importance of interpretability for ethical considerations and the development of trustworthy AI systems \cite{zhang2021survey, samek2021explaining, li2022interpretable}.

\section{White-Box Challenges}

\begin{figure}[t]
  \centering
    \includegraphics[width=0.7\linewidth]{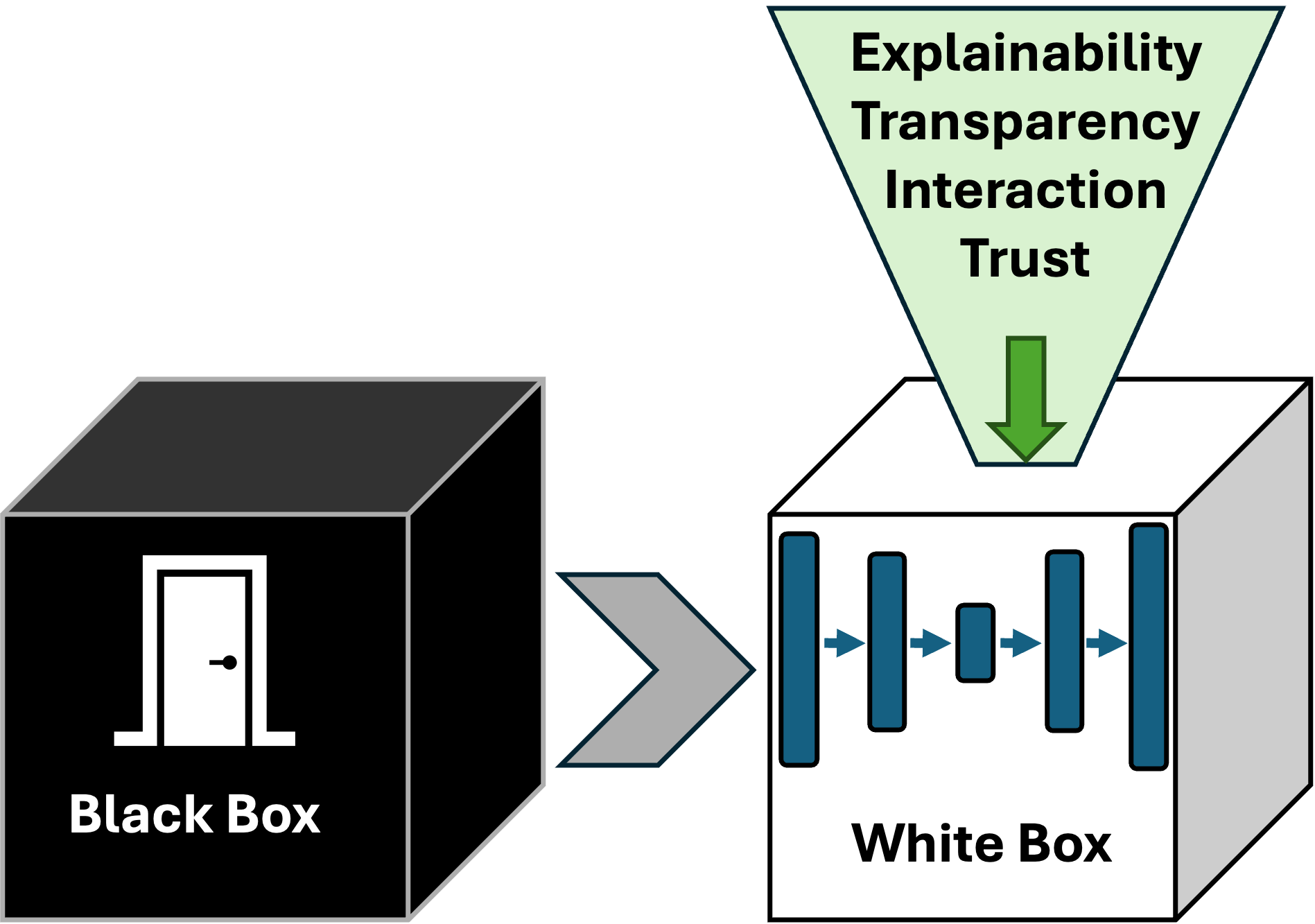}
    \caption{The transformation from Black-Box to White-Box Training by introducing explainability, transparency, interaction, and trust in the development process.}
    \vspace{-10pt}
    \label{fig:box}
\end{figure}

Deep neural networks (DNN) are widely used in HAR and other AI fields due to their ability to capture intricate patterns in time-series data. 
These models can effectively handle the temporal dependencies in HAR data, like transitions between activities, which may span several time steps.
However, while they can achieve high predictive accuracy, black-box models come with significant drawbacks. 
One major issue is the inability to understand or explain why the model makes a certain prediction. 
In HAR, where activities like walking or sitting can overlap or be influenced by sensor noise, the lack of transparency means that practitioners cannot easily diagnose or correct errors in predictions. 
This leads to inefficiencies, as model performance cannot be fine-tuned based on an understanding of its internal workings.

With black-box models, understanding the model’s internal decision-making process is nearly impossible. 
When an activity is misclassified, it is difficult to determine why the model made that mistake, especially when time series input data can not be interpreted intuitively. 
This limits the ability to improve the model, especially when dealing with noisy, imprecise, or incomplete sensor data.
Additionally, sensor artifacts or misplacements can heavily impact the model’s performance. 
Black-box models often mask these errors, leaving users with little insight into what went wrong. 
If the model’s internal layers or neuron weights could be inspected, it might be possible to pinpoint how the model is misinterpreting noisy or poorly calibrated sensor data. 
However, the opacity of these models makes this type of detailed inspection virtually impossible.
The lack of transparency in black-box models further leads to reduced trust in their predictions. 
In sensitive applications like healthcare, where understanding how a model derives its conclusions is crucial, the inability to explain or validate predictions can hinder adoption and acceptance.
Finally, without insight into what the model has learned, improving the model could end in a hit-or-miss process. 
Resources may be wasted on adjustments to the architecture, such as increasing the complexity of the model to properly handle the misclassified sections, when the issue may lie elsewhere—perhaps in the data quality, feature engineering, or simply the weight initialization.

To address the challenges associated with black-box models, the shift to white-box training offers a pathway toward improving interpretability and fostering a deeper understanding of the model’s behavior as shown in \Cref{fig:box}. 
White-box models allow for greater transparency, enabling users to trace and understand the transformations that the data undergoes at each layer of the model. 
For HAR, this transparency is crucial, as it can reveal how sensor data is being processed, how the model’s weights evolve during training, and how the final prediction is formed.
Oppose to designing more transparent model architectures, research must focus on extending existing state-of-the-art architectures towards their explainability in order to maintain the required prediction performance.


\section{Framework}
\begin{figure}[t]
  \centering
    \includegraphics[width=0.8\linewidth]{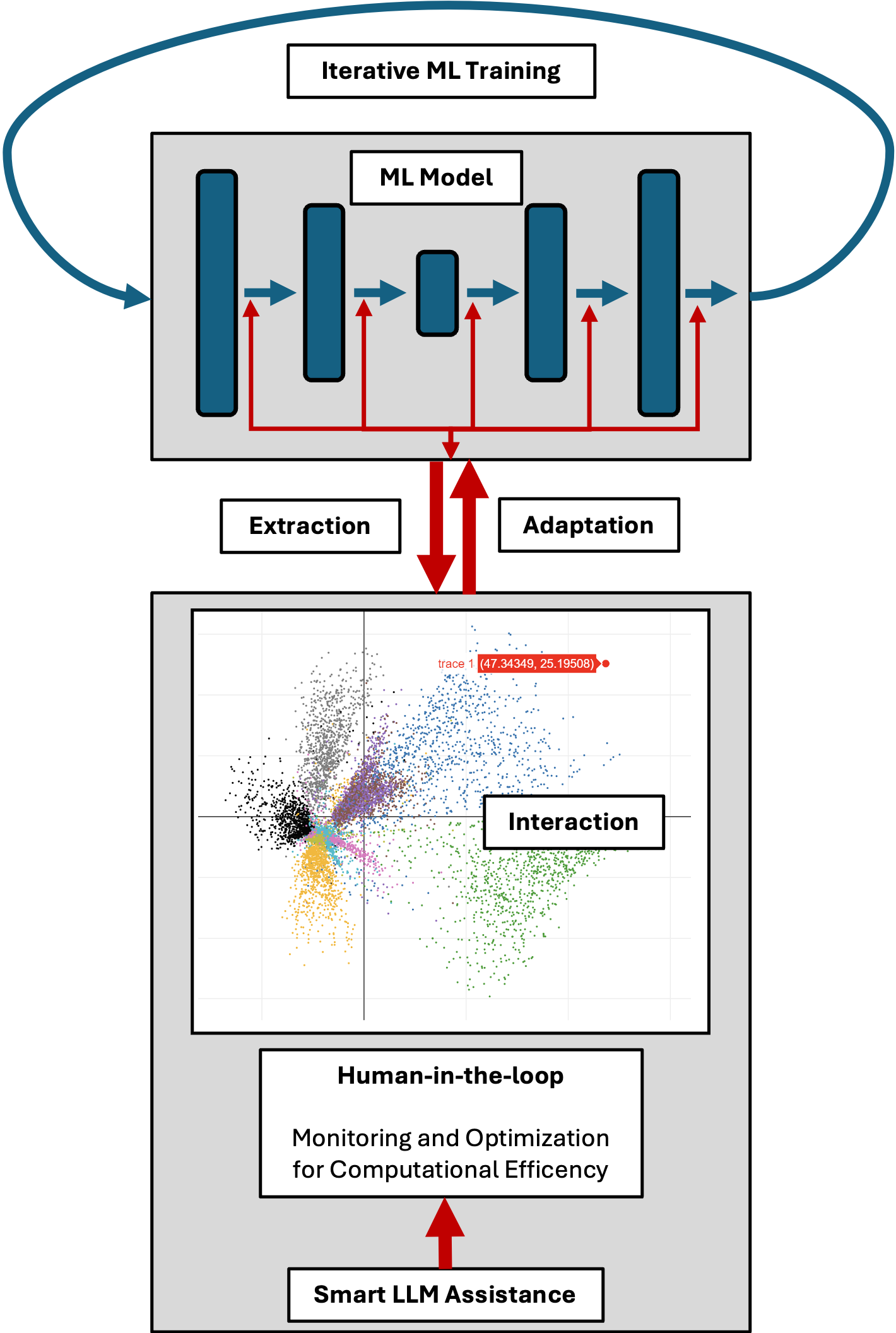}
    \caption{The ML endoscope concept of adding transparency into the traditional Machine Learning training to enhance energy efficiency from a data-driven perspective. }
    \vspace{-10pt}
    \label{fig:concept}
\end{figure}

We envision a framework in the form of an ML Endoscope to enable a deep dive into the intrinsic model behavior from the data-level perspective.
This framework not only facilitates the understanding of how data flows through the network but also supports users in detecting errors, improving model performance, and enhancing training efficiency. 
By providing clear and actionable insights into the model’s behavior, this framework aims to transform how ML developers interact with ML systems, particularly in the context of HAR, where data quality and sensor noise play significant roles in overall performance.
We target to visualize the latent spaces of the model layer outputs to get insights into the data distribution within the model.
We argue that well-structured latent spaces from intermediate model layers support the overall model performance.
The overall concept can be found on \Cref{fig:concept} and will be presented in detail in the following sections.

\subsection{Strategies}

\begin{figure}[!t]
  \centering
  \begin{subfigure}[b]{\linewidth}
    \centering
    \includegraphics[width=0.7\textwidth]{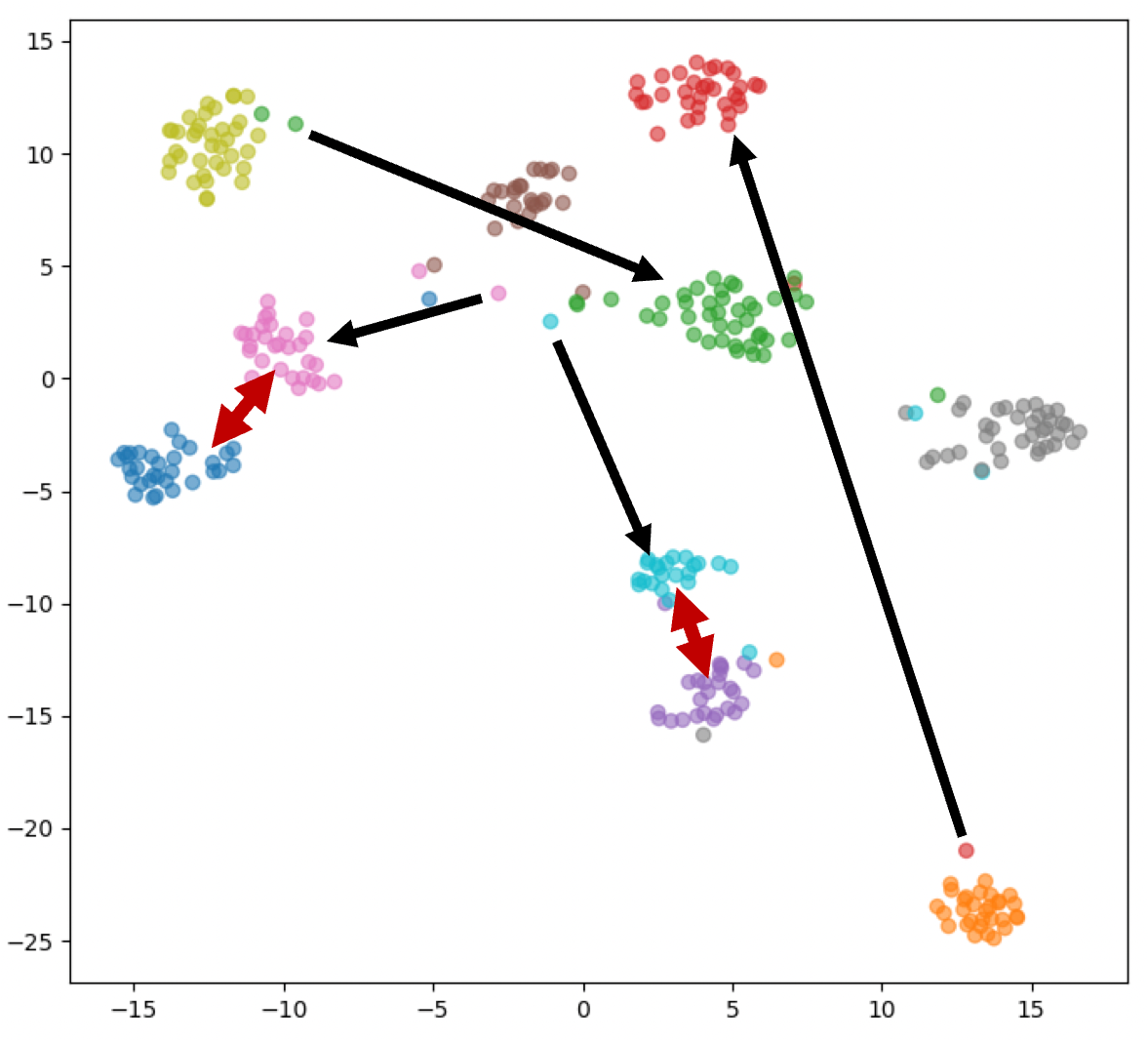}
    \caption{Scatter plot to rejoin outliers with their desired clusters (black arrow) or increase the distance between whole clusters (red arrow).}
    \vspace{10pt}
    \label{fig:scatter}
  \end{subfigure}
  \hfill
  \begin{subfigure}[b]{\linewidth}
    \centering
    \includegraphics[width=0.7\textwidth]{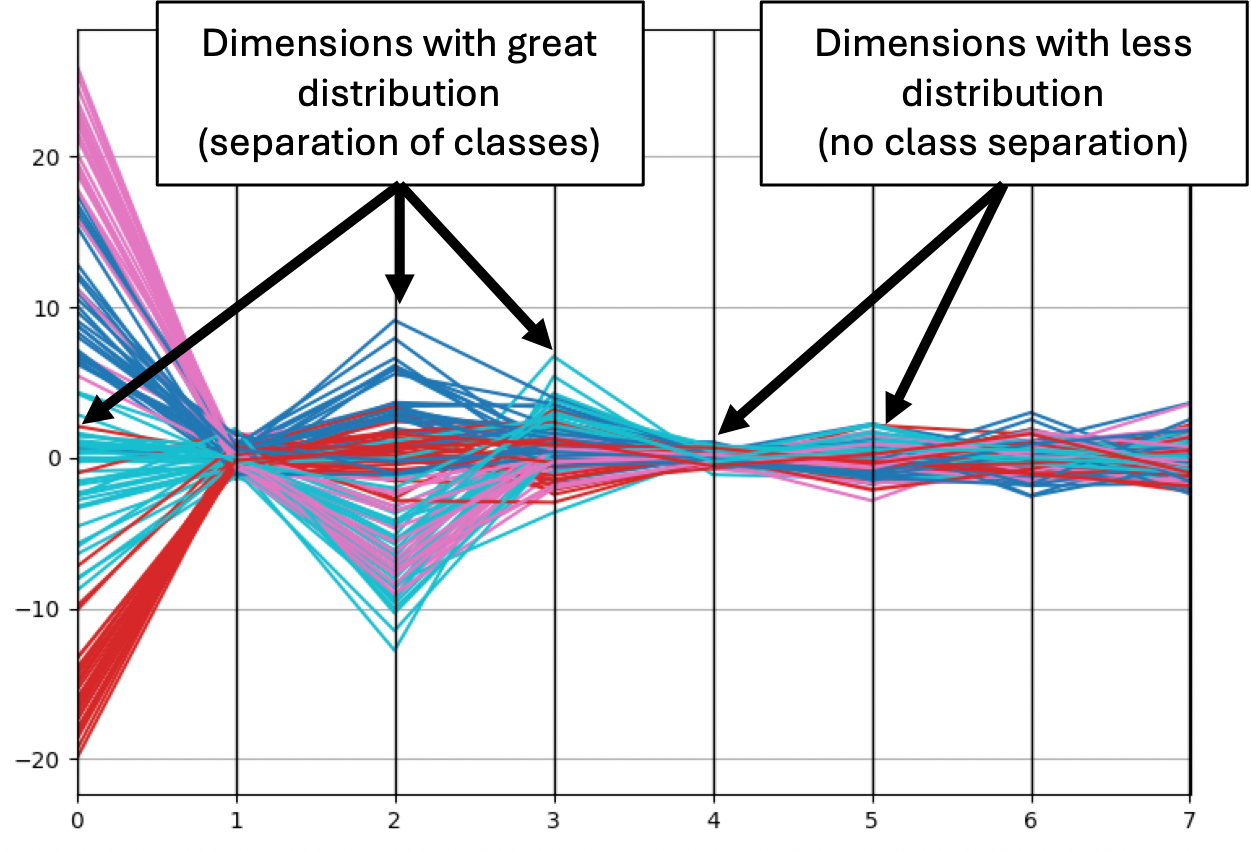}
    \caption{Parallel coordinates plot to investigate the relevance of each dimension.}
    \label{fig:parallel}
  \end{subfigure}
  \hfill
  \begin{subfigure}[b]{\linewidth}
    \centering
    \includegraphics[width=0.7\textwidth]{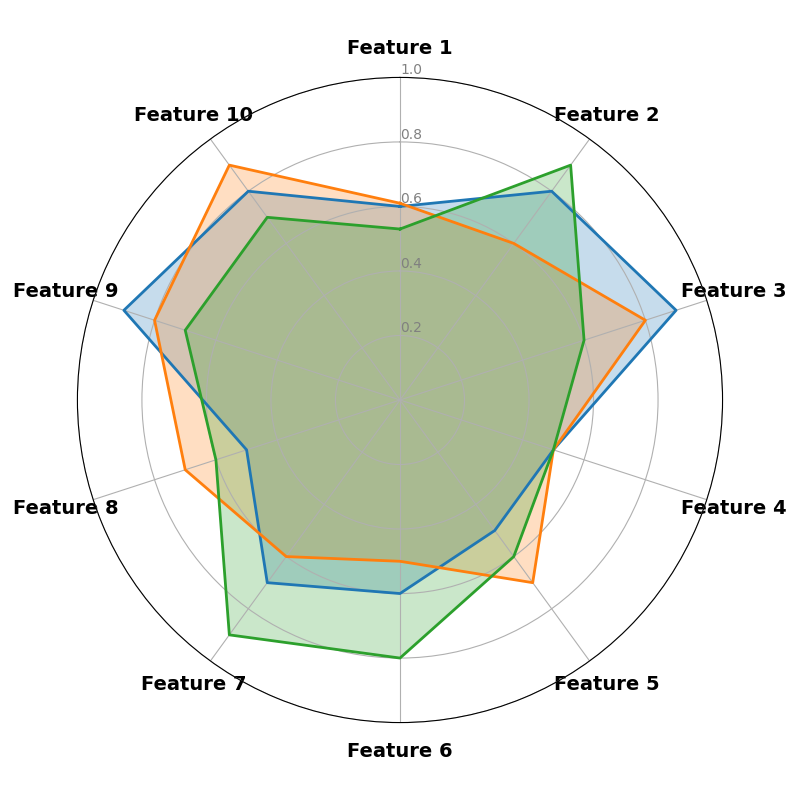}
    \caption{Radar plot to identify dominant features in the latent space.}
    \label{fig:radar}
  \end{subfigure}
  \caption{Latent space visualization strategies to emphasize model and data characteristics.}
  \vspace{-10pt}
  \label{fig:plots}
\end{figure}

Transitioning from black-box to white-box ML models involves adopting several strategies to enhance interpretability without sacrificing performance. 
Various visualization techniques as shown in \Cref{fig:plots}, offer unique perspectives into the latent space dynamics. 
Scatter plots, as shown in \Cref{fig:scatter}, are commonly used to project latent features into two or three dimensions, highlighting clusters that represent distinct activities or behaviors. 
These plots allow users to observe whether the latent space has achieved meaningful separability between classes or whether overlaps exist that could lead to misclassifications. 
Additionally, scatter plots can illustrate outliers that deviate significantly from expected clusters, guiding further investigation into potential anomalies or noisy data points.
Nevertheless, the necessary dimension reduction requires advanced algorithms such as PCA \cite{PCA}, TSNE \cite{van2008visualizing} or UMAP \cite{mcinnes2018umap}, which generate a loss of information due to the reduction of high-dimensional data to human-perceptible 2D or 3D representation.

Oppose to that, parallel coordinate plots are particularly effective for visualizing high-dimensional latent features across different classes. 
Each axis in a parallel coordinates plot corresponds to a specific dimension of the latent space, with lines connecting the values for individual data points. 
This technique enables users to observe relationships and trends across multiple dimensions simultaneously. 
For instance, overlapping lines between classes may indicate features that the model has not effectively differentiated, whereas well-separated lines suggest successful feature extraction.
With that, the relevance of dimensions can be ranked and improved based on the data distribution.

Similarly, Radar plots provide a compact visualization of multidimensional data by mapping latent features onto a circular grid. 
Each axis in the plot represents a latent feature, and the data forms a polygon whose shape varies based on the feature values. 
Radar plots are useful for identifying dominant features in the latent space or for comparing how different activities are represented in terms of their feature profiles.
As shown in \Cref{fig:radar}, data points from different classes can be overlayed and compared.

Temporal dependencies in HAR time-series data can also be explored through dynamic latent space visualizations. 
Animations or continuous tracking of training progressions can illustrate how latent representations evolve, offering a visual narrative of how the model transitions between activities or handles overlapping sequences. 
This temporal perspective helps in identifying transitions or states that the model struggles to distinguish, enabling targeted refinements.

Choosing the right visualization depends on the specific challenges of the dataset and the interpretability needs of the user. 
While scatter plots provide an intuitive overview of separability, parallel coordinates excel at capturing high-dimensional relationships, and radar plots highlight feature-level insights. 
Combining these methods offers a comprehensive understanding of the latent space, revealing where the model excels and where adjustments are needed.

\subsection{Human-In-The-Loop Training}
As an extension of the plain visualization techniques, the proposed ML Endoscope from \Cref{fig:concept} offers an interactive way to incorporate human expertise into the model development process. 
Human-in-the-loop (HITL) allows users to directly interact with the training process by observing and manipulating the model’s internal state, enabling real-time adjustments based on insights from visualizations.

As users interact with the model’s visualizations, they can manually adjust the latent space projections or tweak model parameters when they observe misclassifications or issues with feature separability. 
This immediate feedback loop helps correct errors as they arise, allowing for more effective and efficient model refinement.

Further, HITL also promotes iterative improvements. 
The interactive nature of the framework means that adjustments can be made in real time, and users can immediately assess the impact of their changes on model performance. 
This contrasts with traditional training workflows, where changes often require lengthy retraining processes and a waste of resources. 
The iterative nature of HITL reduces the time spent on trial-and-error adjustments and accelerates the convergence of the model.

Lastly, HITL enhances model interpretability and trust. 
By providing insights into how the model makes decisions and allowing users to trace those decisions, HITL fosters a deeper understanding of the model’s behavior. 
In critical applications such as healthcare, where understanding the reasoning behind predictions is crucial, HITL ensures that the model’s decisions are not only accurate but also transparent.

\subsection{LLM Agent Support}

LLMs are capable of directly processing and understanding high-dimensional data, making them particularly suited for interpreting latent space representations that may be challenging for users to analyze on their own. 
Integrating LLMs into the ML Endoscope may enhance user support by offering real-time, context-aware assistance in interpreting complex visualizations and high-dimensional data. 
Through natural language explanations, the LLM can translate complex patterns and clusters into actionable insights, helping users understand the significance of specific areas in the latent space, such as poorly separated classes or anomalies.

Additionally, LLMs can assist users by analyzing visualizations as presented in \Cref{fig:plots} with the desired metric calculations and observations. 
By combining their language generation capabilities with vision models, such as CLIP \cite{radford2021learning}, LLMs can provide interpretations of plots by explaining what each visualization reveals about the model’s behavior.
For instance, the LLM can highlight overlapping clusters in scatter plots, suggesting that the model may struggle with distinguishing certain classes, or it can point out the key features that drive separability in parallel coordinate plots. 
This combination of LLM and vision model capabilities ensures that users, regardless of their expertise level, receive meaningful, tailored feedback that enhances their understanding of the model and accelerates the optimization process.

However, the efficiency gains in training need to be significant enough to justify the computational costs of frequent LLM interactions, making it essential to balance interpretability benefits with resource usage.
For environments where efficiency is targetted or even limited, strategies such as periodic LLM queries or hybrid approaches that combine LLM insights with simpler, more computationally efficient models could mitigate these resource constraints.

\section{Evaluation}
The evaluation of the proposed framework is designed to comprehensively assess its effectiveness in improving HAR through the introduced white-box and HITL approaches. 
The evaluation combines quantitative metrics for model performance and efficiency, and latent space quality with qualitative insights from domain experts to provide a holistic understanding of the framework’s impact.
Considering the approaches presented in \Cref{sec:related}, the evaluation can be extended with relevant state-of-the-art approaches.

The first aspect of the evaluation focuses on quantifying model performance. 
Using established HAR datasets such as the popular PAMAP2 \cite{reiss2012introducing} dataset, which provides diverse activity data and sensor modalities, the model’s classification accuracy and robustness are analyzed under real-world conditions. 
Metrics such as accuracy, precision, recall, and F1-score can be calculated for each activity class to evaluate consistent performance across categories. 
The proposed white-box approach is compared to traditional black-box models to determine the impact of latent space visualizations and HITL feedback. 
For that, we keep the black-box model architecture while extracting the latent information from the model during the training and feeding potential adjustments back into the layer.
Particular attention is given to how these techniques reduce misclassifications in challenging cases, such as overlapping activities or ambiguous data patterns.
Further, the efficiency enhancement can be correlated with the reduced training duration of white-box training due to faster convergence.

The quality of latent space representations, a foundational element of the framework, is assessed using a range of metrics. 
The well-established silhouette score can be used to quantify the degree of separation between clusters in the latent space, with higher scores indicating well-defined activity classes. 
Complementary metrics such as the Davies-Bouldin Index and the Calinski-Harabasz Score are employed to measure cluster compactness and separation. 
Temporal dynamics of the latent space are also analyzed, visualizing how feature clusters evolve and stabilize over training epochs. 
These quantitative measures are supported by qualitative insights derived from the previously introduced visualizations. 
Together, these evaluations provide a comprehensive view of the latent space’s quality and its role in improving model interpretability and decision boundaries.

The human-centric aspect of the evaluation involves qualitative feedback from domain experts who interact with the HITL framework and white-box visualizations. 
Experts assess the usability and informativeness of latent space visualizations, focusing on how effectively they reveal insights into model behavior and support interventions to correct misclassifications or refine data quality. 
The iterative refinement process enabled by HITL is evaluated for its ability to accelerate model convergence and enhance overall quality. 
Additionally, the integration of LLMs is assessed based on user satisfaction and the perceived value of their explanations and recommendations. 
Experts provide feedback on whether LLM-guided insights and recommendations simplify the interpretation of complex visualizations and optimize workflows, thereby enhancing the overall utility of the framework.
Nonetheless, the effectiveness of LLM support needs to be weighed up with the efficiency considerations, since training efficiency enhancements may become irrelevant due to the excessive use of the LLM.

Finally, a comprehensive evaluation synthesizes the results from numerical performance metrics, energy efficiency profiles, latent space quality assessments, and expert feedback. 
This holistic analysis highlights the framework’s strengths in improving interpretability, efficiency, and classification accuracy while identifying areas for further refinement. 
By addressing both technical and human-centric aspects, the evaluation ensures the framework is well-suited to the challenges of real-world HAR applications, laying the foundation for future advancements in explainable and efficient ML.

\section{Future Work}
Future work will focus on conducting in-depth evaluations of the proposed user interaction and feedback mechanisms, moving beyond conceptual proposals and discussion to practical implementation and validation. 
This includes systematically testing the Human-in-the-Loop (HITL) workflows with diverse user groups to assess their usability, impact, and efficiency. 
Adaptive interfaces will be developed and evaluated to ensure they provide real-time, actionable insights that cater to varying expertise levels, from domain experts to non-technical users. 
The goal is to validate whether these interfaces effectively enable users to manipulate latent space visualizations to enhance model performance.

A critical aspect will involve designing and performing user studies to quantify the effectiveness of the interaction processes. 
These studies will measure how user feedback improves model interpretability, convergence time, and classification accuracy. 
Metrics such as user task completion time, perceived ease of use, and trust in the system will be collected to evaluate the practical usability of the framework. 
Additionally, iterative refinements based on real-world feedback will ensure that the proposed tools address user needs comprehensively.
Additionally, the integration of LLMs into the framework will also be evaluated with a strong focus on how LLM-generated guidance influences user decisions, improves visualization interpretability, and streamlines model optimization.

By focusing on a thorough evaluation of HAR scenarios from popular datasets, we hope to establish the proposed framework as a user-centric solution for explainable AI.

\section{Conclusion}
In this work, we highlighted the challenges of black-box models in HAR, particularly their lack of interpretability and transparency, associated with sensor noise, placement, and inefficiencies in addressing misclassifications. 
We emphasized the need for strategies like visualizing latent spaces and incorporating relevant metrics and human knowledge through HITL approaches. 
By integrating interpretability and user-driven adjustments into the model development process, this approach offers a pathway yet to be explored towards effective, trustworthy, and transparent HAR solutions.



\section*{Acknowledgment}
This work is supported by the European Union’s Horizon Europe research and innovation program (HORIZON-CL4-2021-HUMAN-01) through the "SustainML" project (grant agreement no. 101070408).






\bibliographystyle{IEEEtran}
\bibliography{ref} 

\end{document}